   \documentclass[preprint,showpacs,preprintnumbers,amsmath,amssymb,superscriptaddress]{revtex4}
\usepackage{graphicx} 
\usepackage{dcolumn}  
\usepackage{bm}       

\begin{document}

\preprint{}

\title{Coulomb effects in isobaric cold fission from reactions\ $^{233}$U($n_{\rm{th}}$,f), $^{235}$U($n_{\rm{th}}$,f), ${}^{239}$Pu($n_{\rm{th}}$,f) and $^{252}$Cf(sf)}

\author{M. Montoya}
\email{mmontoya@ipen.gob.pe}

\affiliation{Instituto Peruano de Energ\'ia Nuclear, Canad\'a 1470, San Borja, Lima, Per\'u.}
\affiliation{Facultad de Ciencias, Universidad Nacional de Ingenier\'ia, Av. T\'upac Amaru 210, R\'imac, Lima, Per\'u.}

\date{\today} 

\begin{abstract}
The Coulomb effect hypothesis, formerly used to interpret fluctuations in the curve of maximal total kinetic energy as a function of light fragment mass in reactions $^{233}$U($n_{\rm{th}}$,f), $^{235}$U($n_{\rm{th}}$,f) and $^{239}$Pu($n_{\rm{th}}$,f), is confirmed in high kinetic energy as well as in low excitation energy windows, respectively. Data from reactions $^{233}$U($n_{\rm{th}}$,f), $^{235}$U($n_{\rm{th}}$,f), ${}^{239}$Pu($n_{\rm{th}}$,f) and  $^{252}$Cf(sf) show that, between two isobaric fragmentations with similar $Q$-values, the more asymmetric charge split reaches the higher value of total kinetic energy. Moreover, in isobaric charge splits with different $Q$-values, similar preference for asymmetrical fragmentations is observed in low excitation energy windows.%
\end{abstract}


 \maketitle

\section{Introduction}
\begin{figure}
\centering
\includegraphics[width=8cm]{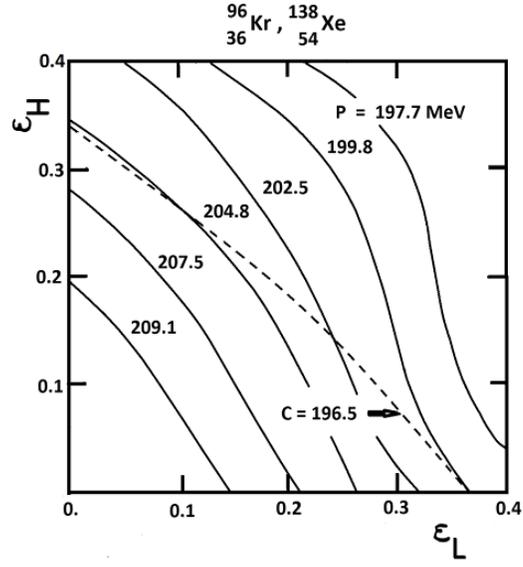}
\caption{Potential energy lines ($P$) equal to Coulomb interaction energy ($C$) plus total deformation energy ($D$), as a function of Nilsson´s deformation parameters of light (${\varepsilon }_{\rm{L}}$) and heavy (${\varepsilon }_{\rm{H}}$) fragments for ${}^{96}_{36}$Kr and $\ {}^{138}_{54}$Xe pair at scission point \cite{Thomas1965}. The scission configuration conformed by the heavy fragment $^{138}$Xe in its spherical ground state, and the fragment $^{96}$Kr in its prolate deformed ground state with an ellipsoidal parameter $\epsilon = 0,4$ gets a potential energy equal to the $Q$-value. The total deformation energy is 3 MeV and the Coulomb interaction energy is 196.5 MeV. Taken from Ref. \cite{Montoya1981}.}
\label{fig:fig1}
\end{figure}

\begin{figure}
\centering
\includegraphics[width=8cm]{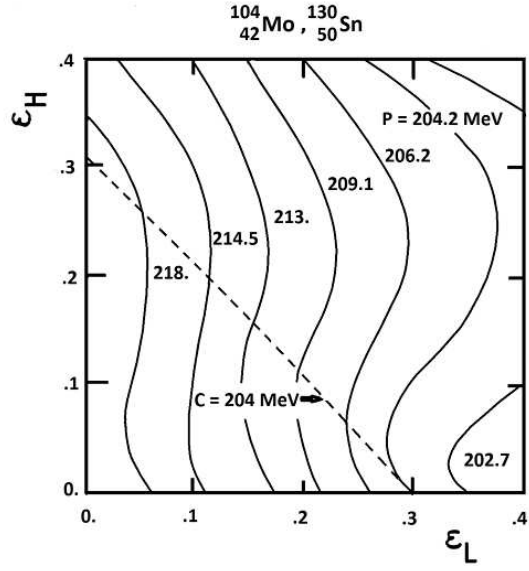}
\caption{ Similar to Fig. ~\ref{fig:fig1}, corresponding to for ${}^{104}_{42}$Mo and $\ {}^{130}_{50}$Sn pair. The scission configuration conformed by the heavy fragment $^{130}$Sn in its spherical ground state, and the fragment $^{104}$Mo in its prolate deformed ground state with an ellipsoidal parameter $\epsilon = 0,3$ gets a potential energy equal to the $Q$-value. The total deformation energy is zero and the Coulomb interaction energy is equal to the $Q$-value. Taken from Ref. \cite{Montoya1981}}
\label{fig:fig2}
\end{figure}

Among the most studied properties of nuclear fission of actinides are the distributions of mass and kinetic energy associated to complementary fragments \cite{Schmitt1966}. Pleasonton found that the highest total kinetic energy is around 190 MeV \cite{Pleasonton1968}. However, those distributions are disturbed by neutron emission. In order to describe one of the consequences of neutron emission, let us suppose that a nucleus with proton number $Z_{\rm{f}}$ and mass number $A_{\rm{f}}$ splits into complementary light (L) and heavy (H) fragments corresponding to primary mass numbers $A_{\rm{L}}$ and $A_{\rm{H}}$, and proton numbers $Z_{\rm{L}}$ and $Z_{\rm{H}}$, having kinetic energies $E_{\rm{L}}$ and $E_{\rm{H}}$, respectively. After neutron emission, those fragments will end with mass numbers
\begin{equation*}
 m_{\rm{L}}=A_{\rm{L}}-n_{\rm{L}}
\end{equation*}
and
\begin{equation*}
 m_{\rm{H}}=A_{\rm{H}}-n_{\rm{H}}
\end{equation*}
where $n_{\rm{L}}$ and $n_{\rm{H}}$  are the numbers of neutrons emitted by the light and heavy fragments, respectively. The corresponding final kinetic energies associated to those fragments will be
\begin{equation*}
e_{\rm{L}} \cong E_{\rm{L}} \left(1- \dfrac{n_{\rm{L}}}{A_{\rm{L}}} \right)
\end{equation*}
and
\begin{equation*}
e_{\rm{H}} \cong E_{\rm{H}} \left(1- \dfrac{n_{\rm{H}}}{A_{\rm{H}}} \right)
\end{equation*}
respectively.  \\\\
In 1979, at the High Flux Reactor (HFR) of Laue-Langevin Institute (ILL), in order to avoid neutron emission effects, the cold fission regions, associated to the highest values of kinetic energy, consequently to the lowest values of total excitation energy, in reactions $^{233}$U($n_{\rm{th}}$,f), $^{235}$U($n_{\rm{th}}$,f) and $^{239}$Pu($n_{\rm{th}}$,f), respectively, were studied by C. Signarbieux {\it et al.} \cite{Signarbieux1981, Montoya1981}. Using the difference of time of flight technique, with solid detectors to measure the fragment kinetic energy, they succeeded to separate neighbouring primary masses, which permitted them to measure the maximal total kinetic energy as a function of primary fragment mass, $K_{\rm{max}}(A_{\rm{L}})$ for light fragment masses from $80$ to $108$. \\\\
In a scission point model was used, assuming that the pre-scission kinetic energy is zero, the maximal total kinetic energy was interpreted as a result of the most compact configuration permitted by the interplay of the Coulomb interaction energy between the two complementary fragments and the total deformation energy limited by the available energy ($Q$-value)  \cite{Signarbieux1981, Montoya1981}. \\\\
The total deformation energy of two complementary fragments ($D$) and the Coulomb interaction energy between these fragments ($C$) are calculated. In the equipotential energy ($P = D + C$) line associated to the $Q$-value, the deformation corresponding to the maximal value of $C$ is selected. In general, the maximal value of $C$ ($C_{\rm{max}}$) is lower than the $Q$-value. For instance, in the reaction $^{233}$U($n_{\rm{th}}$,f), the value of $C_{\rm{max}}$ associated to the fragment pair ($^{96}$Kr, $^{138}$Xe) is 196,5 (3 MeV lower than the corresponding $Q$-value) which is reproduced by a configuration composed by the heavy fragment $^{138}$Xe in its spherical shape and the light fragment $^{96}$Kr in a deformed state with ellipsoidal parameter $\epsilon = 0,4$. See Fig. ~\ref{fig:fig1}.\\\\
The maximal total kinetic energy associated to the mass split $104/130$ reaches the maximal $Q$-value, which corresponds to (${}^{104}$Mo, ${}^{130}$Sn). This kinetic energy is reproduced by the Coulomb interaction energy in a scission configuration conformed by the heavy fragment $^{130}$Sn in its spherical ground state, and the fragment $^{104}$Mo in its prolate deformed ground state with an ellipsoidal parameter $\epsilon = 0,3$. See Fig. ~\ref{fig:fig2}.  Nuclei $^{106}$Mo, $^{107}$Mo and $^{108}$Mo have prolate ground state shapes with ellipsoidal parameter between $0,2$ and $0,3$, but their total kinetic energy values do not not necessarily reach their  corresponding $Q$-values \cite{Montoya1981, Montoya1984, Montoya1986}. The maximal value of total kinetic energy  as a function of mass was interpreted on the base of deformation energy, deformabilities influenced by shell effects and the hypothesis of Coulomb effect \cite{Signarbieux1981, Montoya1981}.\\\\
At the Lohengrin spectrometer in Grenoble, P. Armbruster {\it et al.} \cite{Armbruster1981} measured the mass and kinetic energy as function  of light fragment from reactions $^{233}$U($n_{\rm{th}}$,f) and $^{235}$U($n_{\rm{th}}$,f), respectively. They measured light fragment mass from $80$ to $105$. Like Signarbieux {\it et al.} \cite{Signarbieux1981, Montoya1981} they interpreted the structure of mass yield for cold fragmentation as dominated by potential energy surface of the scission configuration which is influenced by shell effects. They also agree with Signarbieux {\it et al.} that superfluidity is at least partially destroyed. P. Armbruster {\it et al.} conclude that the lowest excitation energy occurs for masses around $92$, while Signarbieux {\it et al.}  \cite{Signarbieux1981} assume that this phenomenum occurs to the fragment pair ($^{104}$Mo, $^{130}$Sn). \\\\
One can easily show that between neighbouring mass fragmentations with similar $Q$-values, the higher maximal Coulomb interaction energy ($C_{\textrm{max}})$ and, consequently, the higher $K_{\rm{max}}$ value, will be reached by the fragmentation with the lower light fragment charge ($Z_{\textrm L}$). This named Coulomb effect reproduces the observed fluctuations in experimental $K_{\rm{max}}(A_{\rm L})$ curves, with a period of 5 units of fragment mass, which is the average of the period of change in the fragment even charge that maximizes the $Q$-value \cite{Montoya1981, Montoya1984, Montoya1986}.\\\\
The mentioned Coulomb effect hypothesis was deduced from the behaviour of the total kinetic energy as a function of fragment mass. In this paper, after a very simple theoretical model explaining the origin of the Coulomb effect, experimental results on charge yield in cold fragmentations, obtained by other authors, that confirm this effect, are presented.
\section{Basics of Coulomb effects in cold fission}
\begin{figure}
\centering
\includegraphics[width=8cm]{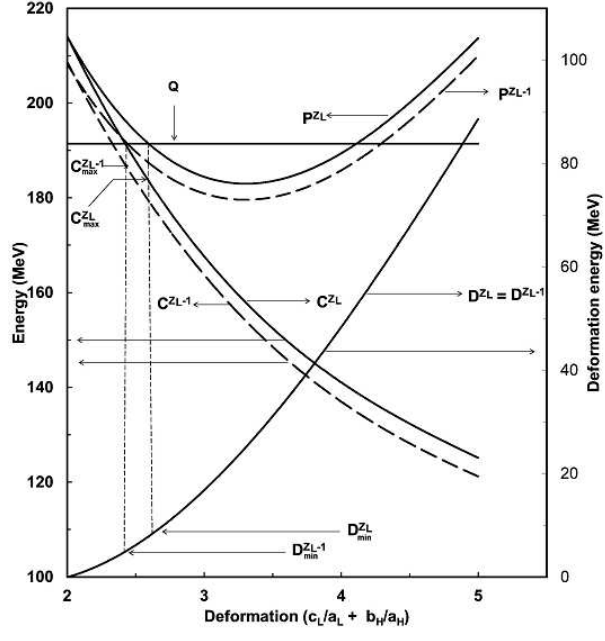}
\caption{Solid lines represent total deformation energy ($D$), Coulomb interaction energy ($C$) and potential energy  ($P=D+C$) of scission configurations as a function of deformation, respectively, for the fragmentation ${}^{96}_{38}{{\rm{Kr}}}/{}^{140}_{54}{{\rm{Xe}}}$. The space of deformation is limited by the total available energy ($Q$). Dashed lines represent similar curves corresponding to the neighbouring more asymmetrical charge split ($38-1)/(54+1$) having the same $Q$-value, for which the minimal deformation is lower than the one corresponding to the first fragmentation (from Ref. \cite{Montoya1986}). As a result the maximal Coulomb interaction energy, which will be converted in kinetic energy, will correspond to the more asymmetrical charge split.}
\label{fig:fig3}
\end{figure}

In a scission point model, the potential energy ($P$) of a scission configuration corresponding to a light fragment charge $Z_{\rm{L}}$ is given by the relation
\begin{equation*}
P^{Z_{\rm{L}}}\left({\mathcal D}\right)=D^{Z_{\rm{L}}}\left({\mathcal D}\right)+C^{Z_{\rm{L}}}\left({\mathcal D}\right)
\end{equation*}
where $D$ is the total deformation energy of fragments, $C$ is the Coulomb interaction energy between complementary fragments, and ${\mathcal D}$ represents the deformed configuration shape. See Fig. ~\ref{fig:fig3}.\\\\
In general, at scission, the fragments have a free energy ($E_{\rm free}$) which is spend in pre-scission kinetic and intrinsec energy of fragments, respectively, obeying the relation
\begin{equation*}
Q = P + E_{\rm free}.
\end{equation*}
The most compact configuration obeys the relation $E_{\rm free}=0$, then
\begin{equation*}
Q = P.
\end{equation*}
In order to explain the Coulomb effect in isobaric splits in cold fission, it matters first to show that, for the same shape configurations, the more asymmetric charge split has a lower Coulomb interaction energy. Let's take the case of two spherical fragments. The Coulomb interaction energy between two complementary hypothetical spherical fragments at scission configuration is given by
\begin{equation*}
C^{Z_{\rm{L}}}_{\rm{sph}}\ =\ \frac{1}{4\pi {\varepsilon }_0}·\frac{Z_{\rm{L}}(Z_{\rm{f}}-Z_{\rm{L}})e^2}{R_{\rm{L}}+R_{\rm{H}}+d}
\end{equation*}
where ${\varepsilon }_0$ is the electrical permittivity, $e$ is the electron charge, $R_{\rm{L}}$ and $R_{\rm{H}}$ are the radii of light and heavy fragment, respectively, and $d$ is the distance between surfaces of fragments. In this paper it is assumed that $d=2$ fm. The nucleus radius for each fragment is given by the relation $R=1.24A^{1/3}$ fm. Then, one can show that
\begin{equation*}
\Delta C_{\rm{sph}}\left(Z_{\rm{L}},Z_{\rm{L}}-1\right)=C^{Z_{\rm{L}}}_{\rm{sph}}-C^{Z_{\rm{L}}-1}_{{\rm{sph}}}=\frac{(Z_{\rm{f}}-2Z_{\rm{L}}+1)}{Z_{\rm{L}}(Z_{\rm{f}}-Z_{\rm{L}})}C^{Z_{\rm{L}}}_{{\rm{sph}}}.
\end{equation*} 
Let us take two cases of charge splits from fission of nucleus $^{236}$U which has $Z_{\rm{F}}=92$. The first case corresponding to $Z_{\rm{L}}=46$ for which the relative variation $\Delta C_{\rm{sph}}$ produced by changing to $Z_{\rm{L}}-1=45$ will be nearly zero; and the second case, a much asymmetric charge split, corresponding to $Z_L=30$, for which the variation $\Delta C_{\rm{sph}}$ produced by changing to $Z_{\rm{L}}-1=29$ will be approximately 3.5 MeV. This gives an idea of how much the Coulomb effect increases with asymmetry of charge split.\\\\
In general, the Coulomb interaction energy between spherical fragments is higher than the $Q-$value. Therefore, in a scission configuration, fragments must be deformed. Let us assume that, for isobaric split $A_{\rm{L}}$/$A_{\rm{H}}$, $C^{Z_{\rm{L}}}({\mathcal D})$ is the interaction Coulomb energy between the two complementary fragments corresponding to light charge $Z_{\rm{L}}$ and scission configuration shape ${\mathcal D}$, with fragments nearly spherical. If one takes two isobaric splits with light fragment charges $Z_{\rm{L}}$ and $Z_{\rm{L}}-1$, respectively, one obtains the relation\\\\
\begin{equation*}
C^{Z_{\rm{L}}}\left({\mathcal D}\right)-C^{Z_{\rm{L}}-1}\left({\mathcal D}\right)\cong \frac{Z_{\rm{f}}-2Z_{\rm{L}}+1}{Z_{\rm{L}}\left(Z_{\rm{f}}-Z_{\rm{L}}\right)}C^{Z_{\rm{L}}}({\mathcal D}).
\end{equation*}
From this relation, for the same shape of scission configuration, one can show that\\\\
\begin{equation*}
C^{Z_{\rm{L}}-1}\left({\mathcal D}\right)<C^{Z_{\rm{L}}}\left({\mathcal D}\right).
\end{equation*}
In consequence, if one assumes that
\begin{equation*}
D^{Z_{\rm{L}}-1}\left({\mathcal D}\right)=D^{Z_{\rm{L}}}\left({\mathcal D}\right)
\end{equation*}
one can show that
\begin{equation*}
P^{Z_{\rm{L}}-1}\left({\mathcal D}\right)<P^{Z_{\rm{L}}}\left({\mathcal D}\right).
\end{equation*}
See Fig. ~\ref{fig:fig3}.
\section{The maximal value of total kinetic energy}
The fragment deformation energy and Coulomb interaction energy between fragments are limited by the $Q$-value of the reaction.  The maximal Coulomb interaction energy corresponding to $Z_{\rm{L}}$ ($C^{Z_{\rm{L}}}_{\rm{max}})$  and the minimal value of deformation energy $D^{Z_{\rm{L}}}_{\rm{min}}$ obeys the relation\\\\
\begin{equation*}
C^{Z_{\rm{L}}}_{\rm{max}}=Q-D^{Z_{\rm{L}}}_{\rm{min}}.
\end{equation*}
Similarly, the relation corresponding to fragmentation with light fragment charge $Z_{\rm{L}}-1$ will be
\begin{equation*}
C^{Z_{\rm{L}}-1}_{max}=Q-D^{Z_{\rm{L}}-1}_{\rm{min}}.
\end{equation*}
The deformation energy ($D$) increases with ${\mathcal D}$. Then the most compact configuration corresponding to $Z_{\rm{L}}-1$ has a lower deformation than the corresponding to $Z_{\rm{L}}$. See Fig. ~\ref{fig:fig3}. In consequence:
\begin{equation*}
D^{Z_{\rm{L}}-1}_{\rm{min}}<D^{Z_{\rm{L}}}_{\rm{min}}
\end{equation*}
From these three relations one deduces that
\begin{equation*}
C^{Z_{\rm{L}}-1}_{\rm{max}}>C^{Z_{\rm{L}}}_{\rm{max}}.
\end{equation*}
Therefore, it is expected that among isobaric splits having similar $Q$- values, the more asymmetric charge  split will reach a more compact configuration, which corresponds to a lower deformation energy, a higher Coulomb interaction energy and, therefore, a higher maximal total kinetic energy. 
\section{Experimental data confirming the Coulomb effect hypothesis}
\begin{figure}
\centering
\includegraphics[width=8cm]{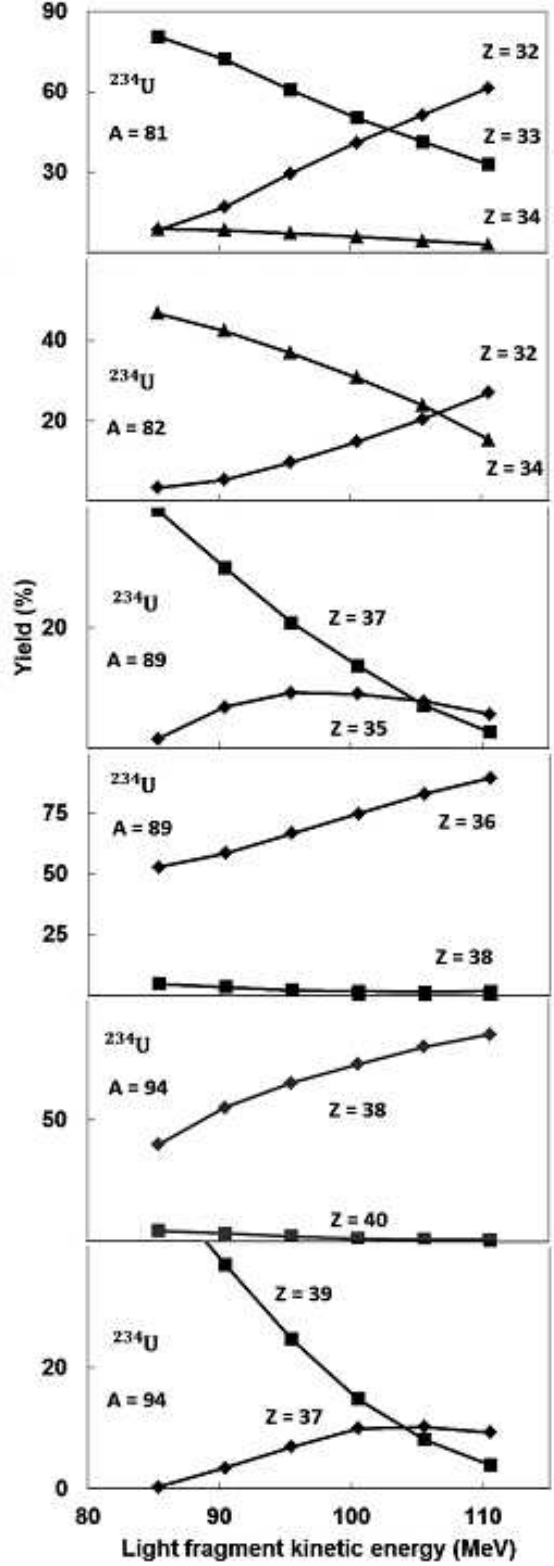}
\caption{Experimental yield of charge, as a function of light fragment kinetic energy, corresponding to some isobaric splits with similar $Q$-values from reaction  $^{233}$U($n_{\rm{th}}$,f), as measured by U. Quade {\it et al.} \cite{Quade1988}.  If one compares yields of odd or even charges, respectively, at the highest measured kinetic energy ($E_{\rm{L}}\ =\ 110.55$ MeV), the higher yield corresponds to the lower light fragment charge.}
\label{fig:fig4}
\end{figure}

\begin{figure}
\centering
\includegraphics[width=8cm]{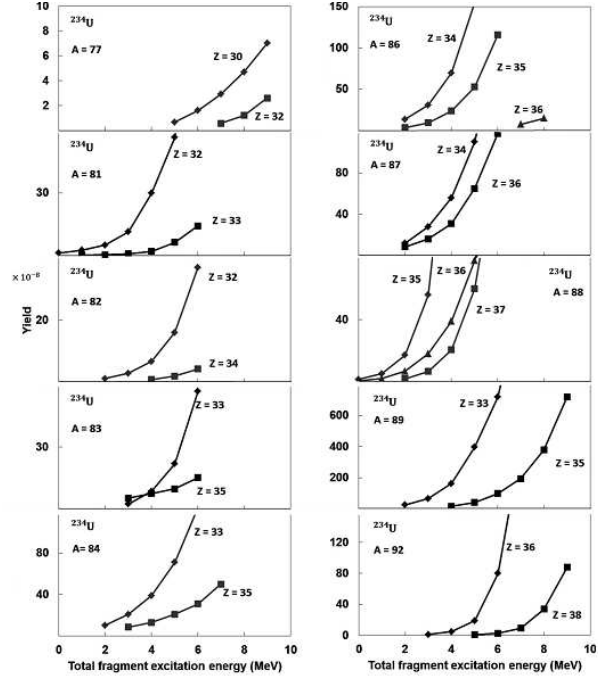}
\caption{Experimental yield of charge, as a function of total excitation energy, from reaction  $^{233}$U($n_{\rm{th}}$,f), as measured by W. Schwab {\it et al.} \cite{Schwab1994}.  If one compares yields of odd or even charges, respectively, at the lowest excitation energy, the higher yield corresponds to the lower light fragment charge.}
\label{fig:fig5}
\end{figure}

\begin{figure}
\centering
\includegraphics[width=8cm]{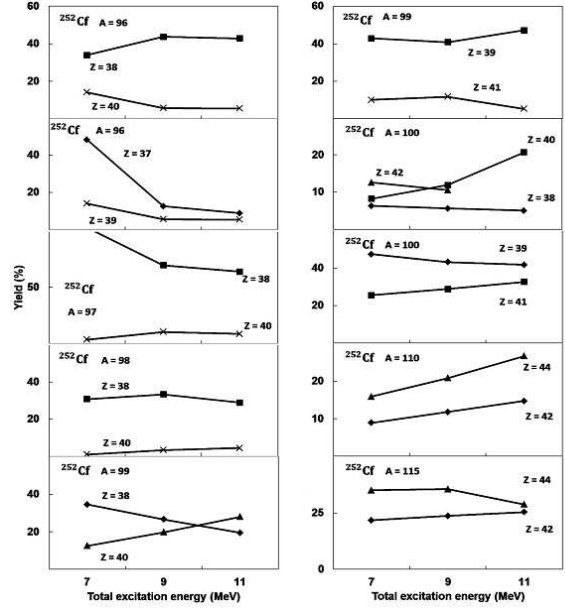}
\caption{Experimental yield of charge, corresponding to total excitation energy values of 7, 9 and 11 MeV, respectively, from reaction  $^{252}$Cf(sf), as measured by F.-J. Hambsch {\it et al.} \cite{Hambsch1993}.  If one compares yields of odd or even charges, respectively, at the lowest excitation energy, the higher yield corresponds to the lower light fragment charge. Exception is observed for mass $100$ corresponding to transitional nuclei, whose deformabilities depends of neutron number.}
\label{fig:fig6}
\end{figure}

According to the Coulomb hypothesis, if the total kinetic energy is due to Coulomb interaction,  in the asymmetrical fragmentation region (light fragment mass lower than 100), it is expected that
\begin{equation*}
K^{Z_{\rm{L}}-1}_{\rm{max}}>K^{Z_{\rm{L}}}_{\rm{max}}
\end{equation*}
Therefore, the higher yield will correspond to the more asymmetrical charge split. Data confirming the Coulomb effect hypothesis will be shown in the following paragraphs. In order to exclude pairing and shell effects in the test of Coulomb hypothesis, one must only take into account charges with same parity and regions exempt of shell transitions .\\\\
In 1984, Clerc {\it et al.} \cite{Clerc1986} measured the charge and mass yield in light fragment kinetic energy windows in reactions $^{233}$U($n_{\rm{th}}$,f) and $^{235}$U($n_{\rm{th}}$,f). The highest kinetic energies studied are 110.55 MeV for $^{233}$U($n_{\rm{th}}$,f) \cite{Quade1988} and 108.0 MeV for $^{235}$U($n_{\rm{th}}$,f) \cite{Armbruster1981}, which correspond to total kinetic energies appreciably below the highest Q-values. They observe that the excitation energy of the fragments increases with increasing asymmetry of the mass split,  result which agrees with the Coulomb effect hypothesis. In order to not exceed  Q-values at least one fragment must be deformed, and that means deformation energy, excepts if the needed deformation corresponds to fragment ground state, as in the case of $^{104}$Mo,  observed by Montoya {\it et al.} \cite{Montoya1981, Montoya1984, Montoya1986}. However, Clerc {\it et al.} interpret this result playing with the distance between fragments centers (R$_c$) corresponding to a Coulomb energy equal to the $Q$-value, the interaction radius R$_{int}$ and tunneling through the barrier separating the “fission valley” from the valley corresponding to two separated fragments (“fusion valley”). If one assumes that those variables have the same values for neighbouring mass and even or odd charge fragmentations, respectively, the Coulomb effect hypothesis still valid in the Clerc {\it et al.}' approach. \\\\
In 1985, Trochon {\it et al.} \cite{Trochon1985, Trochon1989} presented the  mass and charge yield corresponding to the highest values of $K_{\rm{max}}(A_{\rm{L}})$ ($K\ge 194$ MeV) referred to reaction $^{235}$U($n_{\rm{th}}$,f). In this region, the masses $102$ and $104$ correspond to charges $40$ and $42$, respectively. Although the $Q$-value corresponding to charge $40$ is $3$ MeV lower than the corresponding to charge $42$, their corresponding yields are similar. \\\\
Neighbouring masses with different even or odd charges, respectively, with similar $Q$-values can be studied to test the Coulomb effect hypothesis. The fragment with mass $86$ and charge $34$ has a $Q$-value about $2$ MeV lower than the corresponding to the fragment with  mass $88$ and charge $36$ but its  $K_{\rm{max}}$ is $1$ MeV higher.\\\\
Moreover, for the mass split $104/132$ Trochon {\it et al.} observed that only the odd charge split $41/51$ reaches a ``true'' cold fission ($K_{\rm{max}}\cong Q$), while $K_{\rm{max}}$ referred to the magic charge split $42/50$  is ${\rm 3}$ MeV lower than the corresponding $Q$-value \cite{Trochon1989}.\\\\
They also observed that, for isobaric fragmentation, the highest kinetic energy is reached by the charge corresponding to the highest $Q$-value; except for the mass $91$ for which the charge that maximizes $Q$ is $37$, but the highest $K$ is reached by the charge $36$. \\\\
In 1986, G. Simon {\it et al.} \cite{Simon1986} presented results on charge and mass distribution for cold fragmentation from reactions $^{233}$U($n_{\rm{th}}$,f), $^{235}$U($n_{\rm{th}}$,f) and $^{239}$Pu($n_{\rm{th}}$,f), respectively. Among other they arrive to the following conclusions: i) the maximal total kinetic energy are mostly reached by the charges maximizing the $Q$-value ii) for similar $Q$-values the more asymmetric charge split reach higher values of total kinetic energy. \\\\
According to Coulomb effect, a bump in  $K_{\rm{max}}$ as a function of mass must be produced with a change of the charge that maximizes the $Q$-value, which occurs with a period of $5$ or $6$ amu. For the reaction $^{235}$U($n_{\rm{th}}$,f) the change of the light fragment charge that maximizes the $Q$-value occurs for masses $86$, $90$, $96$ and $102$, corresponding to charges $34$, $36$, $38$ and $40$, respectively. In 1991 C. Signarbieux {\it et al.} \cite{Signarbieux1991} showed that the total excitation energy present minimal values for those fragments. \\\\
In 1988 U. Quade {\it et al.} \cite{Quade1988} studied cold fragmentation in the reaction $^{233}$U($n_{\rm{th}}$,f). They measured the charge yield for isobaric splits as a function of light fragment kinetic energy. Some of their results are presented on Fig. ~\ref{fig:fig4}. \\\\
For asymmetrical mass split ${\rm 81/153}$, although $Z_{\rm{L}} =32$ corresponds to a $Q$-value approximately 2 MeV lower than the corresponding to $Z_{\rm{L}} =33$, its probability is higher in the coldest fission region.\\\\
For the mass split $82/152$, although $Z_{\rm{L}} =32$ corresponds to a $Q$-value approximately 4 MeV lower than the corresponding to $Z_{\rm{L}} =34$, its probability is higher in the coldest region. \\\\
For the mass split ${\rm 89/145}$, comparing two odd charge splits referred to $Z_{\rm{L}}=35\ $and $Z_{\rm{L}}=37$, respectively, although $Z_{\rm{L}}=35\ $corresponds to a $Q$-value $1$ MeV lower than the corresponding to $Z_{\rm{L}}=37$, its probability is higher in the coldest region.\\\\ 
Similarly, for the mass split ${\rm 94/140}$, between the two odd charge splits referred to $Z_{\rm{L}}=37$ and $Z_{\rm{L}}=39$, respectively, having a similar $Q$-value, in the cold fission region, the probability for $Z_{\rm{L}}=37$ is higher than the corresponding to $Z_{\rm{L}}=39$. \\\\
U. Quade {\it et al.} noticed, in cold fragmentations, the preferential formation of the element with the highest $Q$-value. However, among the 28 masses they found 10 exceptions. In these exceptions, the highest probability corresponds to a light fragment charge lower that the corresponding to the highest $Q$-value. In Tab. \ref{tbl:tab1} are showed their corresponding masses, the charges that maximize the isobaric $Q$-value, and the charge corresponding to the higher yield at the kinetic energy equal to 110.5 MeV. These results agree with the Coulomb hypothesis: the highest yields do not correspond to the charge that maximize the $Q$-value but to a lower one. \\\\
In 1991 G{\"o}nnenwein {\it et al.} propose the ``Tip model of cold fission'' \cite{Gonnenwein1991} which is a more elaborated version of the model proposed in 1984 by Clerc {\it et al.} \cite{Clerc1986}. They include the deformation properties of nuclei in their ground states. They propose the concept of ``true cold fission'' which correspond to a critical minimum tip distance. This distance, as derived from the theoretical deformations, is assumed to be 3.0 fm.\\\\
In 1994, W. Schwab {\it et al.} show that, for in the reaction $^{233}$U($n_{\rm{th}}$,f), definitely there is a clear trend to prefer more asymmetric charge split in cold fission \cite{Schwab1994}. They calculated the mass and charge yied as a function of excitation energy. Comparing cold isobaric fragmentations with charge with the same parity, in the region of low excitation energy, the higher yield correspond to the lower light fragment charge. See Fig. ~\ref{fig:fig5}.\\\\
In 1993 F.-J. Hambsch {\it et al.} \cite{Hambsch1993} presented experimental data corresponding to spontaneous fission of $^{252}$Cf. They presented the charge and mass yield referred to total excitation energy values equal to 7, 9 and 11 MeV, respectively.  Taking into account the cases with three available excitation energies, corresponding to  cold isobaric fragmentations with charge with the same parity, one can observe that the higher yield corresponds to the lower light fragment charge. See Fig. ~\ref{fig:fig6}. \\\\
The Coulomb effect is more evident in the region associated to the more asymmetric fragmentations. For light fragments heavier than 100 amu, other effects seems to be reflected on charge and mass yield. In the region corresponding to transitional fragments with mass number around 100 and neutron numbers $N \ge 58$ it is expected that deformabilities are reflected on the charge and mass yield. For instance, in the mass fragmentation $100/152$ with 7 MeV of excitation energy, the charge $Z=42$ is preferred to $Z= 38, 40$. In mass fragmentations $110/142$ the yield of $Z = 44$ ($N = 66$) is higher than the corresponding to $Z = 42$ ($N = 68$). For the mass fragmentation $115/137$, the yield of charge $Z = 44$ ($N = 61$) is higher than the corresponding to $Z = 42$ ($N = 63$). 
\section{Conclusion}
It was shown that, in the cold region of thermal neutron induced fission of ${}^{233}$U, ${}^{235}$U, ${}^{239}$Pu and spontaneous fission of $^{252}$Cf, respectively, between isobaric charge fragmentations in the asymmetric region ($A < 100$), with similar $Q$-values of the reaction, the more asymmetric charge fragmentation reaches the higher maximal total kinetic energy. This results is interpreted, in a scission point model, as a ``Coulomb effect'' \cite{Montoya1981, Montoya1984, Montoya1986}: between charge splits with similar $Q$-value, a lower light fragment charge corresponds to a lower Coulomb repulsion, which will permit to reach a more compact configuration and, as a consequence, a lower minimal deformation energy, and a higher maximal Coulomb interaction energy. The final result of that will be a higher maximal fragment kinetic energy. For charge splits with different $Q$-values, the more asymmetrical charge splits are associated to the lower minimal excitation energy. The most compact configuration could be interpreted in terms of fragments shapes \cite{Montoya1981, Montoya1984, Montoya1986} or in terms of distance between fragments \cite{Gonnenwein1991}: more compact configuration is equivalent to lower distance between complementary fragments.
%

\begin{thebibliography}{widest-label}
\bibitem{Schmitt1966} H. W. Schmitt, J.H. Neiler and F. J. Walter, Phys. Rev. \textbf{141} (1966)1146 
\bibitem{Pleasonton1968} F. Pleasonton, Phys. Rev. \textbf{174}, 1500, (1968). 
\bibitem{Brissot1979} R. Brissot, J.P. Bocquet, C. Ristori, J. Crançon, C.R. Guet, H.A. Nifenecker  and M. Montoya: \textit{Proc. of a Symp. on Phys. and Chem. of Fission}, IAEA. Vienna, (1979).
\bibitem{Signarbieux1981} C. Signarbieux, M. Montoya, M. Ribrag, C. Mazur, C. Guet, P. Perrin, and M. Maurel: J. Physique Lettres, \textbf{42}, 437, (1981). 
\bibitem{Montoya1981} M. Montoya: Thesis, Université Paris XI, Orsay, (1981).
\bibitem{Montoya1984} M. Montoya: Z. Phys. A -- Atoms and Nuclei, \textbf{319}, 219-225, (1984). 
\bibitem{Montoya1986} M. Montoya, R.W. Hasse and P. Koczon: Z. Phys. A -- Atoms and Nuclei \textbf{325}, 357-362, (1986). 
\bibitem{Thomas1965} T.D. Thomas, W.M. Gibson and J.G. Safford, \textit{Proc. of Symp. of Physics and Chemistry of Fission}, IAEA. Vienna, (1965).
\bibitem{Quade1988} U. Quade, K. Rudolph, S. Skorka, P. Armbruster, H.-G. Clerc, W. Lang, M. Mutterer, C. Schmitt, J.P. Theobald, F. G{\"o}nnenwein, J. Pannicke, H. Schrader, G. Siegert, D. Engelhardt: Nucl. Phys. \textbf{A487}, 1-36, (1988). 
\bibitem{Wapstra1985} A.H. Wapstra and G. Audi, Nucl. Phys. \textbf{A432}, 1, (1985).
\bibitem{Schwab1994} W. Schwab, H.-G. Clerc, M. Mutterer, J.P. Theobald and H. Faust: Nucl. Phys. \textbf{A577}, 674-690, (1994).
\bibitem{Trochon1985} J. Trochon, G. Simon, J.W. Behrens, F. Brisard and C. Signarbieux: \textit{Proc. Int. Conf. on Nuclear Data for Basic and Applied Science}, 13-17 May 1985, Santa Fe, New Mexico, Radiation Effects \textbf{92}, 327 - 332, (1986)
\bibitem{Trochon1989} J. Trochon {\it et al.}: \textit{Proc. Conf. ``50 Years with Nuclear Fission``}, Am . Nucl. Soc., 313, (1989).
\bibitem{Armbruster1981} P. Armbruster, U. Quade, K. Rudolph, H.-G. Clerc, M. Mutterer, J. Pannicke, C. Schmitt, J.P. Theobald, D. Engelhardt, F. G{\"o}nnenwein and H. Schrader The cold fragmentation of 234 U in 233 U(n th ,f) 4th International Conference on Nuclei far from Stability, Helsingør, proceedings CERN 81-09, Geneva (1981)
\bibitem{Simon1986} G. Simon, J. Trochon and C. Signarbieux, Fission meeting, Arcachon (France), 14-17 Oct 1986, CEA-CONF— 8860.
\bibitem{Signarbieux1991} C. Signarbieux, 1st Intern. Conf. on Dynamical Aspects of Nuclear Fission,. Smolenice, Slovakia, J. Kristiak and B.I.Pustylnik, eds., J.I.N.R., Dubna, 1992, p. 19.
\bibitem{Clerc1986} H.-G. Clerc, W. Lang, M. Mutterer, C. Schmitt, J.P. Theobald, U. Quade, K. Rudolph, P. Armbruster, F. G{\"o}nnenwein, H. Schrader and D. Engelhardt, Nucl. Phys. \textbf{A552}, 277 - 295, (1981)
\bibitem{Gonnenwein1991} F. G{\"o}nnenwein and B. Börig:  Nucl. Phys. \textbf{A530}, 27-57, (1985).
\bibitem{Hambsch1993} F.-J. Hambsch, H.-H. Knitter and C. Budtz-Jorgensen Nuclear Physics \textbf{A554},  209-222, ( 1993).
\end{thebibliography}

%
\pagebreak
\begin{table}
 \centering
 \caption{${}^{233}$U(n${}_{th}$,f). The ten light fragment masses for which the highest probabilities, at kinetic energy of 110.5 MeV, correspond to charges lower than the referred to the highest $Q$-value.  This table is based on information from Ref. \cite{Quade1988}.}
\centering

\begin{tabular}{rrr}  
A & Z of highest  & Z of highest  \\
  & $Q$-value &  yield \\  
82 & 34 & 33 \\ 
86  & 36  & 34 \\ 
82  & 34  & 33 \\ 
87  & 36  & 35 \\ 
92  & 38  & 37 \\ 
102  & 42  & 40 \\ 
103  & 42  & 41 \\ 
85  & 34,35  & 34 \\  
81  & 33  & 32 \\ 
91  & 37  & 36 \\ 
101  & 41  & 40 \\ 
\label{tbl:tab1}
\end{tabular}
\end{table}
\end{document}